\documentclass[useAMS,usenatbib]{mn2e}
\usepackage[]{graphicx}
\def\figspath{figs}
\def\AInS{\mbox{$A_{\rm In}^{\odot}$}}
\def\Am{\mbox{$A_{\rm In}^{\rm{m}}$}}
\def\InI{\hbox{In\,{\sc i}}}
\def\InII{\hbox{In\,{\sc ii}}}
\def\ScI{\hbox{Sc\,{\sc i}}}
\def\ScII{\hbox{Sc\,{\sc ii}}}
\def\SmI{\hbox{Sm\,{\sc i}}}
\def\SmII{\hbox{Sm\,{\sc ii}}}
\def\VI{\hbox{V\,{\sc i}}}
\def\FeI{\hbox{Fe\,{\sc i}}}
\def\FeII{\hbox{Fe\,{\sc ii}}}
\def\TiI{\hbox{Ti\,{\sc i}}}
\def\CrI{\hbox{Cr\,{\sc i}}}
\def\RuI{\hbox{Ru\,{\sc i}}}
\def\TaI{\hbox{Ta\,{\sc i}}}
\def\gf{\mbox{$g \! f$}}

\title{On the solar abundance of indium}
\author[N. Vitas et al.]
  {N.~Vitas,$^{1,2}$\thanks{E-mail: N.Vitas@astro.uu.nl}
  I.~Vince,$^{2,3}$ M.~Lugaro,$^{1,4}$ O.~Andriyenko,$^{5,6}$
  M.~Go\v{s}i\'{c},$^2$ and R.~J.~Rutten$^{1, 7}$
  \newauthor \\
  $^1$Sterrekundig Instituut, Utrecht University, P.O.~Box 80000, 3508 TA Utrecht, The Netherlands\\
  $^2$Department of Astronomy, University of Belgrade, Studenstki trg 16, 11 000 Belgrade, Serbia\\
  $^3$Astronomical Observatory, Volgina 7, 11160 Belgrade, Serbia\\
  $^4$Center for Stellar \& Planetary Astrophysics, P.O.~Box 28M
Monash University, Victoria, 3800, Australia\\
  $^5$ICAMER, NASU, 27 Akademika Zabolotnoho St., 03680 Kyiv, Ukraine\\
  $^6$Main Astronomical Observatory, NASU, 27 Akademika Zabolotnoho St., 03680 Kyiv, Ukraine\\
  $^7$Institutt for Teoretisk Astrofysikk, University of Oslo, P.O.~Box 1029 Blindern, N-0315 Oslo, Norway\\
  }

\begin{document}

\date{Accepted 2007 .
      Received 2007 September ?;
      in original form 2007 September ?}

\pagerange{\pageref{firstpage}--\pageref{lastpage}} \pubyear{2007}

\maketitle

\label{firstpage}

\begin{abstract}The generally adopted value for the solar abundance of indium
is over six times higher than the meteoritic value. We address this
discrepancy through numerical synthesis of the 451.13~nm line on
which all indium abundance studies are based, both for the quiet-sun
and the sunspot umbra spectrum, employing standard atmosphere models
and accounting for hyperfine structure and Zeeman splitting in
detail. The results, as well as a re-appraisal of indium
nucleosynthesis, suggest that the solar indium abundance is close to
the meteoritic value, and that some unidentified ion line causes the
451.13~nm feature in the quiet-sun spectrum.

\end{abstract}


\section{Introduction}\label{sec:intro}

The solar abundance of indium is controversial because its generally
accepted value significantly exceeds the meteoritic value. At a
factor of six difference, this remains the largest unexplained
discrepancy between meteoritic and solar abundance values. In this
paper we address this problem by considering the nucleosynthesis of
indium and through indium line synthesis for the quiet solar
photosphere and sunspot umbrae including hyperfine structure in
detail.

The meteoritic indium abundance is \mbox{$\Am = 0.80 \pm 0.03$}
(\citealt{2003ApJ...591.1220L} and references therein) where $A_{\rm
In} \equiv \log(n_{\rm In}/n_{\rm H}) + 12$ with $n_{\rm In}$ and
$n_{\rm H}$ the indium and hydrogen particle densities,
respectively. Table~\ref{tab:abundances} summarises the
determinations of the solar indium abundance in the literature. All
measurements are based on a single, very weak feature in the
quiet-sun spectrum at $\lambda = 451.1307$~nm which is commonly
identified as one of the \InI\ resonance lines. The initial result
of \citet{1960ApJS....5....1G} was based on an erroneous oscillator
strength. The other three determinations scatter around $\AInS =
1.6$, the value listed in the compilation of
\citet{2005ASPC..336...25A}. The 0.8~dex discrepancy with the
meteoritic value cannot be explained by the usual uncertainties of
abundance determination such as line strength measurement, imprecise
atomic data and solar modeling deficiencies.

\begin{table}
\caption{$\AInS$ determinations} \label{tab:abundances}
\centering
\begin{tabular}{@{}lccc}
\hline
   Authors                      &  $\AInS$ & Specified error \\
\hline
   \citet{1960ApJS....5....1G}  & 1.16       &       \\
   \citet{1969MNRAS.142...71L}  & 1.71       &       \\
   \citet{1998SSRv...85..161G}  & 1.66       & 0.15  \\
   \citet{2002SoPh..211....3B}  & 1.56       & 0.20  \\
\hline
\end{tabular}
\end{table}

The origin of elements heavier than \hbox{Fe} is mostly attributed
to neutron-capture processes \citep[see][for a
review]{1994ARA&A..32..153M}. $Slow$ neutron capture (the $s$
process) occurs for relatively low neutron densities ($\simeq 10^7
$~cm$^{-3}$), while $rapid$ neutron capture (the $r$ process) occurs
for relatively high neutron densities ($> 10^{20}$~cm$^{-3}$).
\citet{1978ApJ...219..307A} analysed the Sn/In abundance ratio. He
found that `no combination of $r$- or $s$-process products even
remotely resembling those which generally predict the solar system
abundances very successfully can give Sn/In as low as 1.4' (which
results from taking $\AInS = 1.71$).

Recently, \citet{2006MNRAS.370L..90G} suggested that because of its
low condensation temperature (536~K, \citealt{2003ApJ...591.1220L}),
indium may have been depleted in chondritic meteorites to an
abundance much smaller than the solar one. However, we note that,
e.g., thallium has a similar condensation temperature (532~K,
\citealt{2003ApJ...591.1220L}) and a similar meteoritic abundance
($0.78 \pm 0.04$) whereas its well-determined photospheric abundance
($0.9 \pm 0.2.$) is only slightly higher than the meteoritic one. In
a sequel paper \citet{2006MNRAS.371..781G} studied the indium
abundance in a sample of 42 sun-like stars of which five are known
to host planets. He found a strong negative correlation between the
[\hbox{In}/\hbox{Fe}] and [\hbox{Fe}/\hbox{H}] logarithmic abundance
ratios. This trend is much steeper than the comparable relation for
europium, which is a pure $r$-process element. However, one would
expect a less steep trend for indium because it received
contributions both from the $s$- and the $r$-processes.

In this paper we once again scrutinize the solar indium abundance,
paying specific attention to indium nucleosynthesis
(Sect.~\ref{sec:ratio}), indium line identification and appearance
in the solar spectrum (Sect.~\ref{sec:solarindium}), and indium line
synthesis accounting for hyperfine structure both for quiet sun and
sunspot umbrae (Sect.~\ref{sec:synthesis}). The conclusion is that,
after all, the solar indium abundance is likely to be close to the
meteoritic value.

\section{Analysis}\label{sec:analysis}

\subsection{Nucleosynthesis and the S\lowercase{n}/I\lowercase{n}
ratio}\label{sec:ratio}

In this section we update the analysis of
\citet{1978ApJ...219..307A} considering the latest models for the
origin of heavy elements. The indium and tin abundances in the solar
system are believed to have received contributions from both the $s$
and the $r$ processes, with roughly 30 and 60 per cent $s$-process
production to In and Sn, respectively (e.g.
\citealt{2004ApJ...601..864T}, \citealt{1999ApJ...525..886A}).

\begin{table}
\begin{minipage}[t]{\columnwidth}
\caption{Sn/In from different nucleosynthetic processes.}
\label{tab:ratio} \centering
\renewcommand{\footnoterule}{}  
\begin{tabular}{lr}
\hline
 & Sn/In \\
\hline
$s$ process (classical)$^a$ & 39 \\
$s$ process (classical)$^b$ & 46 \\
$s$ process (stellar)$^a$ & 38 \\
$s$ process (GCE)$^c$ & 34 \\
$r$ process (weak)$^d$ & 32 \\
$r$ process (main)$^d$ & 6 \\
solar value $^e$ & 2.51 \\
meteoritic value $^f$ & 20.42 \\
\hline
\end{tabular}
\\
$^a$\citet{1999ApJ...525..886A}, $^b$\citet{2004ApJ...617.1091S},
$^c$\citet{2004ApJ...601..864T}, $^d$\citet{2007ApJ...662...39K},
$^e$\citet{2005ASPC..336...25A}, $^f$\citet{2003ApJ...591.1220L}
\end{minipage}
\end{table}

In Table~\ref{tab:ratio} we present the Sn/In ratios associated to
the most recent models of the $s$ and the $r$ processes. For the
$s$-process we list different estimates of the ratio: two of them
were derived using the classical approach for the $s$ process, where
abundances are calculated via a parametric model, and another via
the stellar model, where the building up of $s$-process abundances
is modeled inside an asymptotic giant branch (AGB) stars (see
\citealt{1999ApJ...525..886A} for details). A further, more
realistic description of the solar system distribution of abundance
is given by models of the galactic chemical evolution (GCE), where
yields from different generations of AGB stars are integrated in
order to build up the solar system abundances at the time and
location of the formation of the sun. In all descriptions the Sn/In
ratios remain very similar among the $s$-process estimates because
during the $s$-process, and far from nuclei with closed neutron
shells, $s$-process abundances are determined by the $\sigma_A N_A
\simeq constant$ rule, where $\sigma_A$ is the neutron capture cross
section of isotope with atomic mass $A$, and $N_A$ its abundance
during the $s$-process. The $\sigma_A$ of In and Sn are determined
with roughly 10 per cent
uncertainties\footnote{http://nuclear-astrophysics.fzk.de/kadonis/}
and thus the Sn/In ratios from the $s$ process have small nuclear
uncertainties.

For the $r$ process we take the recent parametric models of
\citet{2007ApJ...662...39K}. Two $r$-process components, generated
assuming different neutron densities, are introduced to build up the
$r$-process abundances in the solar system: the `weak' $r$ process
produces elements up to tellurium, and the `main' $r$ process
produces elements from tellurium up to the actinides. The Sn/In
ratio for the main $r$ process component is equal to 6
(Table~\ref{tab:ratio}), which confirms the estimate of
\citet{1978ApJ...219..307A} made using the simple argument that the
$r$ process would produce similar yields for close-by isotopes and
that In and Sn have 1 and 6 stable isotopes that can be produced by
the $r$ process, respectively.

Finally, we note that recent $r$-process calculations indicate that
the situation is likely to be much more complex, with several
different components involved \citep{2005farouqi}. However, also in
these more recent models no component with $\mbox{Sn/In} < 8$ is
found (K.-L. Kratz, personal communication).

\subsection{Indium in the solar spectrum.}\label{sec:solarindium}

\begin{figure}
\includegraphics[width=80mm,keepaspectratio]{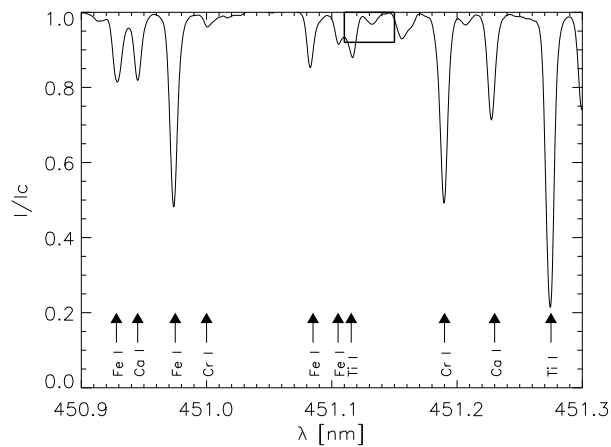}
\caption{The spectral region around 451.13~nm in the Kitt Peak atlas
of the quiet-sun disc-centre photosphere. The intensity scale is
normalised by the adjacent continuum value. The arrows specify line
identifications copied from the Kitt Peak atlas.}
\label{fig:broad_region}
\end{figure}

\begin{figure}
\includegraphics[width=80mm,keepaspectratio]{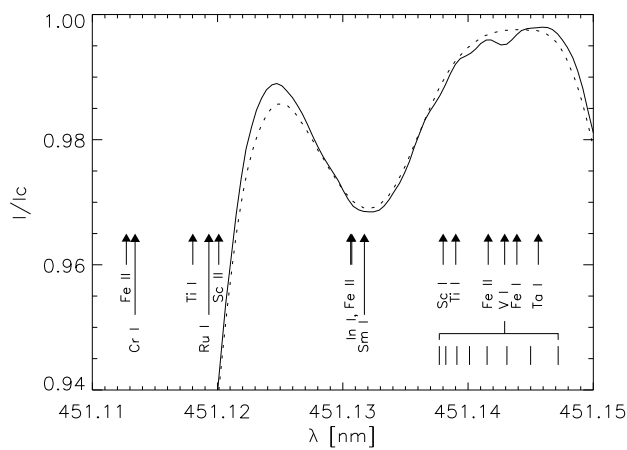}
\caption{Enlargement of the frame in Fig.~\ref{fig:broad_region}.
Solid: Kitt Peak atlas. Dotted: Jungfraujoch atlas. The candidate
lines of Table~\ref{tab:blends} are indicated by arrows. Positions of hyperfine
components of the \VI\ line are also indicated.} \label{fig:narrow_region}
\end{figure}

\paragraph*{The 451.13~nm line.} The resonance line of \InI\ at 451.13~nm
($5p ^2P_{3/2} \rightarrow 6s ^2S_{1/2}$) is the only indium line
that has been identified in the solar spectrum.
Figure~\ref{fig:broad_region} shows this region in the disc-centre
quiet-sun Kitt Peak spectral atlas of \citet{1998assp.book.....W}.
The small frame in Fig.~\ref{fig:broad_region} is enlarged in
Fig.~\ref{fig:narrow_region} and shows the region around 451.13~nm
in detail. There are no significant differences with the
Jungfraujoch atlas \citep{1973apds.book.....D}. In the Kitt Peak
atlas the relatively strong feature at $451.1155$~nm is attributed
to \TiI, but this identification does not agree with
\citet{1995KurCD..23.....K} whose table instead suggests a line of
\CrI\ at 451.1134~nm.

Other lines that may be expected in this narrow region are listed in
Table~\ref{tab:blends} and indicated in
Fig.~\ref{fig:narrow_region}. The lines of \RuI\ and \ScII\ are
fully blended by the wing of the line at $\lambda = 451.1155$~nm.
The four \FeII\ lines have high excitation energy ($> 10$~eV) and
low oscillator strengths, hence they should not be present in the
quiet-sun spectrum. There are three candidate blends of the In line
due to \ScI, \VI\ and \SmI. All three belong to elements with a
large nuclear spin, so that hyperfine splitting must be taken into
account. Scandium and vanadium have only one stable isotope with $I
= 7/2$. Samarium has 7 stable isotopes, two ($^{147}$\hbox{Sm} and
$^{149}$\hbox{Sm}) having nonzero nuclear spin $I = 7/2$.
Unfortunately, data on hyperfine splitting are only available for
the \VI\ line \citep{1995KurCD..23.....K}. The ground level of \SmI\
is split in 7 sub-levels, giving numerous weak transitions as
indicated in Fig.~\ref{fig:narrow_region}. The \SmI~451.1317~nm
line, attributed to \SmII\ by \citet{2006MNRAS.371..781G}, is one of
them. The \InI\ line is not affected by the \TaI~451.1456~nm line as
it is sufficiently far away.

\begin{figure}
\includegraphics[width=80mm,keepaspectratio]{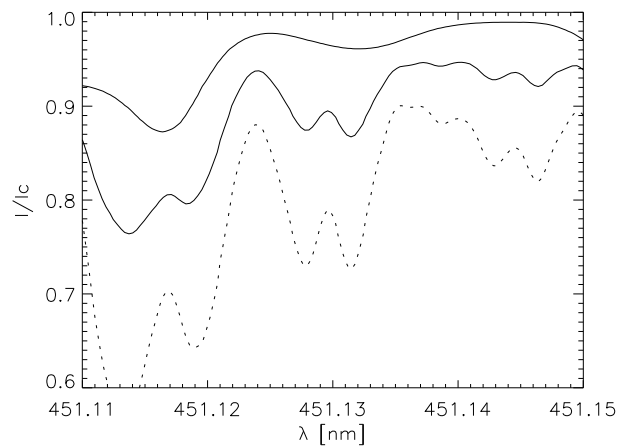}
\caption{The spectral region around the 451.13~nm line in the Kitt
Peak sunspot atlas. Lower solid curve: sunspot spectrum. Dotted:
after subtraction of 60 per cent straylight. Upper solid curve: Kitt
Peak quiet-sun atlas for comparison. Each atlas profile is
normalised to its own adjacent continuum.} \label{fig:kp_sunspot}
\end{figure}

\begin{table}
\caption{Possible blends of \InI~451.13~nm. The data are
from \citet{1995KurCD..23.....K} (K\&B) and \citet{1999A&AS..138..119K} (VALD).}
\label{tab:blends}\centering
\begin{tabular}{l r r c c }
\hline
Element & $\lambda$ [nm] & $E_l$ [eV] & log($\gf$) & Reference\\
\hline
\FeII   & 451.1127    &     11.31 &  $-2.811$ & K\&B \\
\CrI    & 451.1134    &      3.17 &  $-2.366$ & K\&B \\
\TiI    & 451.1170    &      0.50 &  $-3.300$ & VALD \\
\RuI    & 451.1193    &      1.69 &  $-1.070$ & K\&B \\
\ScII   & 451.1201    &      7.87 &  $-1.694$ & K\&B \\
\FeII   & 451.1306    &     10.60 &  $-3.338$ & K\&B \\
\InI    & 451.1307    &      0.27 &  $-0.213$ & K\&B \\
\SmI    & 451.1317    &      0.50 &  $-0.013$ & K\&B \\
\ScI    & 451.1380    &      1.85 &  $-1.193$ & K\&B \\
\TiI    & 451.1390    &      3.29 &  $-4.076$ & VALD \\
\FeII   & 451.1416    &     11.29 &  $-2.063$ & K\&B \\
\VI     & 451.1429    &      2.13 &  $-1.520$ & K\&B \\
\FeII   & 451.1439    &     10.72 &  $-3.405$ & K\&B \\
\TaI    & 451.1456    &      0.70 &  $-1.730$ & K\&B \\
\hline
\end{tabular}
\end{table}

The  umbral spectrum in the Kitt Peak Sunspot Atlas of
\citet{2000asus.book.....W} shows a much stronger line at
\mbox{$\lambda = 451.13$~nm} as illustrated in
Fig.~\ref{fig:kp_sunspot}. Its core is clearly split in two strong
components, with the red peak slightly stronger than the blue one.
\citet{2006MNRAS.371..781G} attributes them to Zeeman splitting of
the \InI\ line. The upper solid curve in Fig.~\ref{fig:kp_sunspot}
represents the photospheric spectrum as a reference. The dotted
curve is the observed umbral spectrum with sizable correction for
straylight following \citet{1965zwaan}, setting its amount to 60 per
cent (i.e., the observed spectrum contains 60 per cent photospheric
and 40 per cent umbral light). The choice of this value is discussed
below. The broad triple-peaked feature to the right of the indium
line is likely due to the hyperfine splitting of the \VI\ line
indicated in Fig.~\ref{fig:narrow_region}.

\paragraph*{Other indium lines.}
Table~\ref{tab:iniiniilines} lists the other candidate lines of
neutral and singly ionised indium. The other component of resonance
multiplet (1) is $5p ^2P_{1/2} \rightarrow 6s ^2S_{1/2}$ at $\lambda
= 410.1765$~nm. It is located close to the centre of the Balmer
H$\delta$ line and is not recognizable in the quiet-sun Kitt Peak
atlas. The other indium lines are absent or completely blended or
located in spectral regions that are not covered by the present
solar atlases.

Since indium is mostly ionised (over 99 per cent throughout the
photosphere assuming the Saha distribution), one would expect that
the \InII\ resonance line at 158.6~nm should be the strongest in the
solar spectrum, but also this line is not clearly present in the
SUMER atlas of \citet{2001A&A...375..591C}.

\begin{table*}
\begin{minipage}{150mm}
\caption{Candidate lines of \InI\ and \InII\
and their appearance in the Kitt Peak atlas (K,
\citealt{1998assp.book.....W}) and the SUMER atlas (S,
\citealt{2001A&A...375..591C}). The wavelengths and oscillator
strengths are from CD-ROM No.~23 of \citet[]{1995KurCD..23.....K}.}
\label{tab:iniiniilines} \centering
\begin{tabular}{c r r c c l}
\hline
Ionisation stage & $\lambda$~[nm]& log($\gf$) & $E_{\rm l}$~[eV]  &  Atlas & Comment \\
\hline
I  &   303.9357 & $-0.143$  &  0.000 &   K & In the wing of \hbox{Fe\,{\sc i}} 303.9321~nm\\
I  &   325.6087 & $ 0.170$  &  0.274 &   K & In the wing of a strong line at $\lambda = 325.613$~nm\\
I  &   325.8559 & $-0.620$  &  0.274 &   K & Between two strong lines\\
I  &   410.1765 & $-0.550$  &  0.000 &   K & In the core of H$\delta$\\
I  &   451.1307 & $-0.213$  &  0.274 &   K & The only identified line of \InI\\
I  &   684.7440 & $-1.200$  &  3.022 &   K & No line\\
I  &   690.0132 & $-1.510$  &  3.022 &   K & Strong telluric line\\
II &    78.3892 & $-3.092$  &  0.000 &   S & No line\\
II &    91.0951 & $-1.777$  &  0.000 &   S & No line\\
II &    92.7324 & $-3.170$  &  0.000 &   S & Heavily blended \\
II &   158.6450 & $ 0.393$  &  0.000 &   S & Very weak unidentified line\\
\hline
\end{tabular}
\end{minipage}
\end{table*}

\subsection{Synthesis of the solar \InI~451.13~nm line}\label{sec:synthesis}
\paragraph*{Model atom.}
Oscillator strengths for the indium transitions were taken from
\citet{1995KurCD..23.....K} except for the 451.13~nm line for which
Table~\ref{tab:loggf} lists various values
from the literature. We adopted the value of \citet{2006crchandbook}. The
indium hyperfine structure must be taken into account because the nuclear
spin of In is $I = 9/2$. Values of the magnetic dipole ($A$) and
electric quadrupole ($B$) constants of the hyperfine interaction for the
lower and upper levels of the \InI\ multiplet 1 were
taken from \citet{1981jackson} and \citet{1978JPhB...11.2821Z}.

\begin{table}
\caption{Values of log($\gf$) for \InI~451.13 nm
transition.}\label{tab:loggf}\centering
\begin{tabular}{l c l} \hline
log($\gf$)  &   $A$ [10$^{8}$~s$^{-1}$] &  Source\\
\hline
$-0.308$   &      0.806  &  L\\
$-0.265$   &      0.890  &  M\\
$-0.213$   &      1.003  &  K\&B\\
$-0.206$   &      1.019  &  F\&W\\
$-0.167$   &      1.115  &  C\&B\\
$-0.360$   &      0.715  &  G\&K W\\
$-0.590$   &      0.421  &  G\&K D\\
\hline
\end{tabular}
\medskip

L = \citet{1969MNRAS.142...71L}, M = \citet{2000ApJS..130..403M},
K\&B = \citet{1995KurCD..23.....K}, F\&W = \citet{2006crchandbook}, C\&B =
\citet{1962etps.book.....C}, G\&K = \citet{1989fsss.book.....G} (W =
value determined from the equivalent width, D = from the line depth)
\end{table}

The level splitting was evaluated from:

$$\Delta E_F = \frac{C}{2} A + \frac{3 C (C+1) - 4 I (I+1) J (J+1)}{8I(2I-1)J(J-1)}B,$$

\noindent where $C = F(F+1) - J(J+1) - I(I+1)$, $F = J+I, J+I-1,
\dots, |J-I|$, and $J$ is the electronic angular momentum
\citep[e.g.,][]{1992sobelman}. For the upper level with $J = 1/2$
only the first member in the Hamiltonian of the hyperfine
interaction is present and is split into two sublevels ($F =$ 4, 5),
while the upper level is split into four sublevels ($F =$ 3, 4, 5,
6). Hence, the selection rule ($\Delta F = -1, 0, 1$) implies that 6
hyperfine components are expected in the 451.13~nm transition.
Relative intensities of these components were calculated from Wigner
$6j$ coefficients assuming analogy with LS coupling
\citep[cf.][]{1992sobelman}. Table~\ref{tab:compos} specifies the
resulting hyperfine structure components. Indium has two stable
isotopes, but since the ratio of their abundances is $^{115}$In :
$^{113}$In = 95.7 : 4.3, the second isotope can be neglected.

\begin{table}
\caption{Hyperfine components of \InI~451.13~nm with $r$ the
intensity of the component relative to the strongest one and
$\Delta\lambda$ the wavelength shift from the centre of
gravity.}\label{tab:compos}\centering
\begin{tabular}{c c r r}
\hline  $F_l$  &  $F_u$  &  $r$     &  $\Delta\lambda$~[pm]\\
\hline
5   &     4   & 33.84   & $ 3.14$\\
4   &     4   & 50.76   & $ 2.38$\\
3   &     4   & 53.84   & $ 1.93$\\
6   &     5   &100.00   & $-1.38$\\
5   &     5   & 50.76   & $-2.58$\\
4   &     5   & 18.46   & $-3.34$\\
\hline
\end{tabular}
\end{table}

\paragraph*{Quiet-sun profile synthesis.}
The radiative transfer code {\sc multi 2.2} of
\cite{1992ASPC...26..499C} was used for detailed synthesis of the
\InI~451.13~nm line profile.

\begin{figure}
\includegraphics[width=80mm,keepaspectratio]{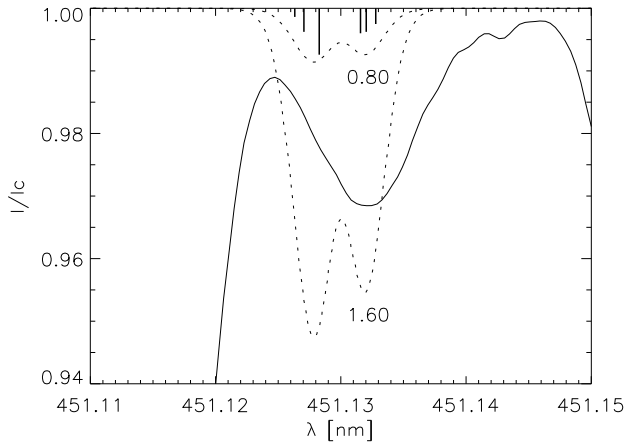}
\caption{Synthetic profiles (dotted) of the \InI~451.13~nm line for $\AInS = 0.80$ and 1.60 compared with the Kitt
Peak atlas (solid). The vertical bars at the top specify the
wavelengths and relative intensities of the hyperfine components.
They produce double peaks in the synthetic profiles.}
\label{fig:synthetic_profile}
\end{figure}

As commonly done in solar abundance analysis
\citep[see][]{2002JAD.....8....8R}, we assume local thermodynamical
equilibrium (LTE) and the HOLMUL model of
\citet{1974SoPh...39...19H} for the solar photosphere, including its
microturbulence stratification. For the oscillator strength of
\InI~451.13~nm we take the value of \citet{2006crchandbook}. We
computed line profiles both for the meteoritic abundance value $\Am
= 0.80$ and for the solar value $\AInS = 1.60$ listed by
\citet{2005ASPC..336...25A}. The resulting profiles are compared
with the observed quiet-sun profile in
Fig.~\ref{fig:synthetic_profile}. The vertical bars specify the
hyperfine structure components. It is obvious that neither
computation fits the observed feature at all.

\begin{figure}
\includegraphics[width=80mm,keepaspectratio]{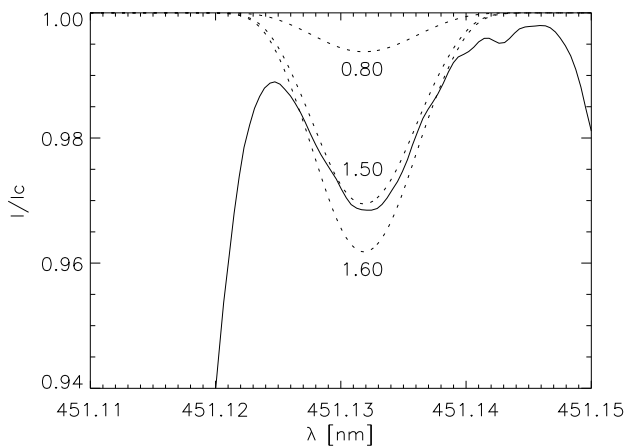}
\caption{Synthetic profiles (dotted) of the \InI~451.13~nm line compared with the Kitt Peak atlas (solid) after ad-hoc
convolution and wavelength shift.} \label{fig:fit}
\end{figure}

Figure~\ref{fig:fit} illustrates the steps that are required to
force a better match. The synthesized line was convoluted with a
broad Gaussian profile with $\mbox{FWHM} = 1.58$~pm $(1~\mbox{pm} =
10^{-3}$~nm), much wider than the instrumental broadening of the FTS
spectrometer at Kitt Peak, and it was shifted redward over $\Delta
\lambda = 2.35$~pm. With these ad-hoc measures a reasonable fit is
obtained for indium abundance $\AInS = 1.50$ (Fig.~\ref{fig:fit})
but the assumptions made to obtain it are not justified. We conclude
that the solar line at this wavelength in the quiet-sun spectrum is
probably not due to indium.

\begin{figure}
\includegraphics[width=80mm,keepaspectratio]{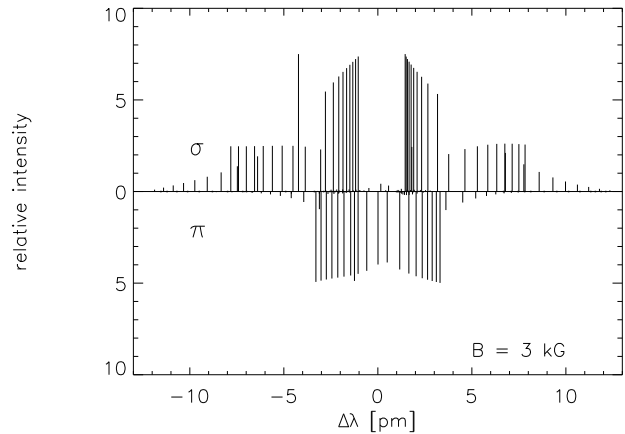}
\caption{Combined Zeeman and hyperfine splitting of the \InI~451.13~nm line a
magnetic field of 3~kG. } \label{fig:hfs_components_in_H}
\end{figure}

\paragraph*{Umbral profile synthesis.}
We use the appearance of the 451.13~nm line in the sunspot spectrum
to provide an independent estimate. In order to include Zeeman
splitting we estimate the magnetic field to be 3000~G from the
\TiI~464.52~nm line observed at the same day as the \InI~451.13~nm
in the Kitt Peak Sunspot atlas. Lines with hyperfine structure split
in complex manner in the presence of magnetic fields
\citep{1975A&A....45..269L}. We employed the code by
\citet{1978A&AS...33..157L}. The result is shown in
Fig.~\ref{fig:hfs_components_in_H}. We assume that the orientation
of the magnetic field in the observed sunspot is purely vertical and
maintains only the $\sigma$ components.

\begin{figure}
\includegraphics[width=80mm,keepaspectratio]{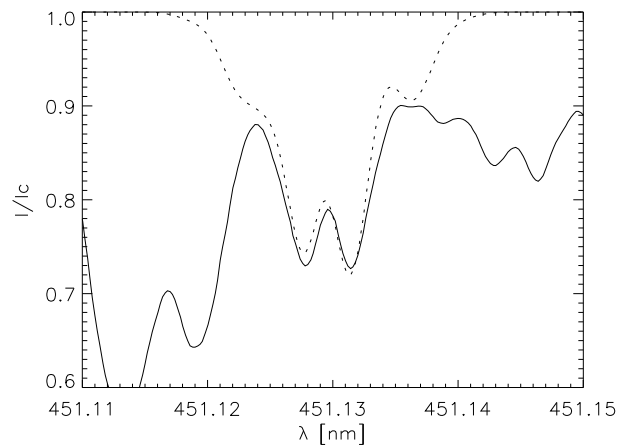}
\caption{Profiles of the \InI~451.13~nm line in the sunspot
spectrum. Solid: Kitt Peak sunspot atlas with 60 per cent straylight
subtraction. Dotted: computed profile with instrumental broadening
corresponding to the Kitt Peak FTS (Gaussian with $\mbox{FWHM} =
0.47$~pm).} \label{fig:sunspot_profile}
\end{figure}

For the model atmosphere we use the semi-empirical umbral model M of
\citet{1986ApJ...306..284M}. Figure~\ref{fig:sunspot_profile}
compares the synthesized profile assuming the meteoritic abundance
value $\Am = 0.80$ to the observed profile assuming 60 per cent
straylight. The dotted curve is the computed profile after
convolution with instrumental broadening corresponding to the FTS
resolution of one million. No ad-hoc wavelength shift is applied.
The match is quite satisfactory, including the separation and
amplitude ratio of the two peaks.

\begin{figure}
\includegraphics[width=80mm,keepaspectratio]{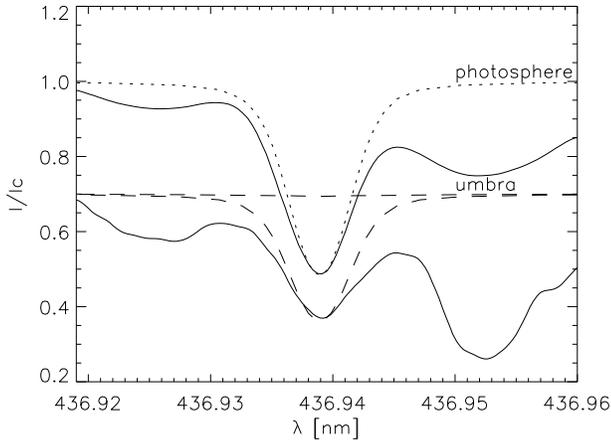}
\caption{Profiles of the \FeII~436.9~nm line in quiet-sun (upper
pair) and umbral spectra (lower curves). The latter are shifted down
over 0.3 of the relative intensity scale. Solid: observed profiles
from the Kitt Peak photosphere and sunspot atlases. Dotted: computed
profile from the HOLMUL model. Dashed: computed profiles using the
Maltby M model without and with 65 per cent photospheric straylight
addition.} \label{fig:iron1}
\end{figure}

\begin{figure}
\includegraphics[width=80mm,keepaspectratio]{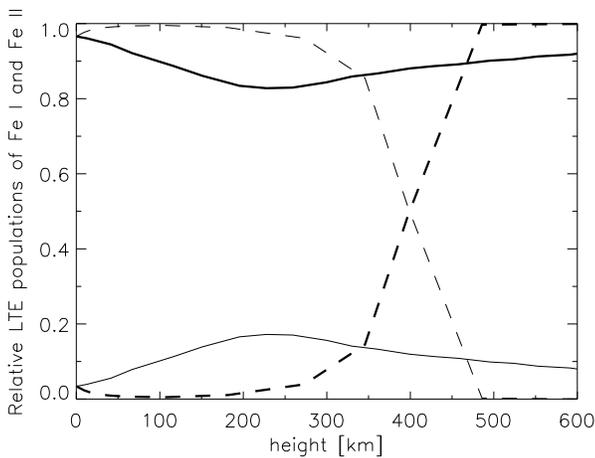}
\caption{The relative LTE populations of neutral (thin curves) and
once-ionised iron (thick) in the HOLMUL photosphere model (solid)
and the Maltby M umbral model (dashed).} \label{fig:iron2}
\end{figure}

Obviously, the good match in Fig~\ref{fig:sunspot_profile} requires
60 per cent straylight subtraction. We first argue and then show
that this is a reasonable value. In their introduction to the Kitt
Peak sunspot atlas, \citet{2000asus.book.....W} already remark that
photospheric straylight fully dominates the violet part of the
spectrum. Its contribution scales as $\lambda^{-2}$ (see
\citealt{1970SoPh...12..328S}) so that it is still large in the blue
around 450~nm. \citet{2003A&A...412..513B} used 25 per cent to fit
their TiO lines at 705~nm which translates into 61 per cent at
450~nm with the $\lambda^{-2}$ dependence. In order to quantify this
estimate more precisely, we have performed spectral synthesis of the
\FeII~436.9~nm line chosen because it is the closest \FeII~line
without Zeeman splitting to the \InI~451.13~nm line. The oscillator
strength of the line was adapted to obtain a good fit to the
photospheric spectrum as shown in Fig.~\ref{fig:iron1}. The lower
curves demonstrate that the \FeII~line vanishes in the intrinsic
umbral spectrum and only appears due to the contribution by
photospheric stray light. The good fit to the atlas spectrum was
obtained with 65 per cent addition of the computed photospheric
spectrum. The reason why the \FeII~line vanishes in the umbral
spectrum is shown in Fig.~\ref{fig:iron2}. The dominant ionisation
stage is \FeII~ in the photosphere, but \FeI~throughout the low
umbral photosphere. \FeII~lines that are weak in the photospheric
spectrum can therefore not appear in the intrinsic umbral spectrum.

\section{Discussion}\label{sec:discussion}

Our update in Sect.~\ref{sec:ratio} of the analysis of
\citet{1978ApJ...219..307A} did not change the conclusion that there
is no combination of nucleosynthetic processes that can produce a
Sn/In ratio as obtained using the most recent solar values, while
the chondritic ratio can be easily accounted for. This is a strong
indication that the solar abundance of indium cannot be much higher
than the meteoritic value. This also suggested by the absence of the
\InII\ resonance line at 158.6~nm.

Furthermore, Fig.~\ref{fig:synthetic_profile} shows that standard
line synthesis including hyperfine structure does not fit the
observed profile for either the high or the low abundance value. In
contrast, its appearance in the sunspot spectrum is well reproduced,
including hyperfine Zeeman splitting, using the low meteoritic
abundance value (Fig.~\ref{fig:sunspot_profile}) and assuming a
reasonable straylight fraction (60 per cent) and umbral field
strength of 3000~G.

These results together suggest that the line at 451.13~nm in the
quiet-sun spectrum is not due to indium but to some other species.
It should be from an ion with low ionisation energy of the neutral
stage since the line disappears in the sunspot spectrum (cf.\
Fig.~\ref{fig:iron2}). The observation by
\citet{2006MNRAS.371..781G} of significant trends of the assigned
indium abundance with $T_{\rm{eff}}$ and [Fe/H] in a sample of stars
cooler than the sun suggests that the unidentified transition should
have high excitation energy. The best possible candidate in the list
of \citet{2006MNRAS.371..781G} would be the \SmII\ line, but as
noted in Sect.~\ref{sec:solarindium} this is actually a resonance
transition of \SmI.

Finally, an alternative explanation of the \InI~451.13~nm line
formation might be NLTE effects. In particular, the appearance of
the 451.13~nm line as emission line in the spectra of long-period
variables during the downward modulation phase was explained already
by \citet{1937ApJ....86..499T} as optical pumping between H$\delta$
and a \InI\ line at 410.1~nm which shares its upper level with
\InI~451.13~nm. However, we this mechanism does not operate at all
in the deep solar photosphere where the formation of the hydrogen
Balmer lines is close to LTE (cf.\ Fig.~30 of
\citealt{1981ApJS...45..635V}).

In conclusion, we suggest that the solar indium abundance is close to the meteoritic value as
evidenced in the sunspot spectrum, whereas the feature at 451.13~nm in the quiet-sun spectrum
remains unidentified and is likely to be an ion line at high excitation energy from a species
with low first ionisation energy.

\section*{Acknowledgments}

We are indebted to Stevan Djeni\v{z}e for drawing our attention to
the solar indium abundance and to Pit S\"utterlin and Karl-Ludwig
Kratz for discussions. We thank the referee for suggesting the
analysis in Fig.~\ref{fig:iron1}. This research project has been
supported by a Marie Curie Early Stage Research Training Fellowship
of the European Community's Sixth Framework Programme under contract
number MEST-CT-2005-020395. ML is supported by NWO (VENI fellow).
The Ministry of Science of Serbia partially supported this research
(project `Stellar and Solar Physics', Contract No.~146003).

\bibliographystyle{mn2e}
\bibliography{Vitas_et_al}

\bsp

\label{lastpage}

\end{document}